\documentclass[aps,prb,reprint,superscriptaddress]{revtex4-2}
\usepackage{graphicx}% Include figure files 
\usepackage{epstopdf}
\usepackage{dcolumn}% Align table columns on decimal point nothing
\usepackage{bm}% bold math not found
\usepackage{amsmath}
\usepackage{hyperref}% add hypertext capabilities
\hypersetup{backref, colorlinks=true, linkcolor=blue, citecolor=blue, urlcolor=blue}
\usepackage{sistyle} % pour angstrom, et unité SI :   
\usepackage{multirow}
\newcolumntype{C}[1]{>{\centering\let\newline\\\arraybackslash\hspace{0pt}}m{#1}}

\begin{document}

\title  {Chiral spin spiral in synthetic antiferromagnets probed by circular dichroism in x-ray resonant magnetic scattering}

\author{Cyril L\'eveill\'e}
\affiliation{Synchrotron SOLEIL, L'Orme des Merisiers, Saint-Aubin, BP 48, 91192 Gif-sur-Yvette Cedex, France}

\author{Samuel Flewett}
\affiliation{Pontifica Universidad Cat\'{o}lica de Valpara\'{i}so, Avenida Universidad 330, Valparaíso, Chile}

\author{Erick Burgos-Parra}
\affiliation{Synchrotron SOLEIL, L'Orme des Merisiers, Saint-Aubin, BP 48, 91192 Gif-sur-Yvette Cedex, France}
\affiliation{Unit\'e Mixte de Physique, CNRS, Thales, Universit\'e Paris-Saclay, 91767, Palaiseau, France}

\author{Yanis Sassi}
\affiliation{Unit\'e Mixte de Physique, CNRS, Thales, Universit\'e Paris-Saclay, 91767, Palaiseau, France}

\author{William Legrand}
\affiliation{Unit\'e Mixte de Physique, CNRS, Thales, Universit\'e Paris-Saclay, 91767, Palaiseau, France}

\author{Fernando Ajejas}
\affiliation{Unit\'e Mixte de Physique, CNRS, Thales, Universit\'e Paris-Saclay, 91767, Palaiseau, France}

\author{Vincent Cros}
\affiliation{Unit\'e Mixte de Physique, CNRS, Thales, Universit\'e Paris-Saclay, 91767, Palaiseau, France}

\author{Nicolas Reyren}
\affiliation{Unit\'e Mixte de Physique, CNRS, Thales, Universit\'e Paris-Saclay, 91767, Palaiseau, France}

\author{Nicolas Jaouen}
\affiliation{Synchrotron SOLEIL, L'Orme des Merisiers, Saint-Aubin, BP 48, 91192 Gif-sur-Yvette Cedex, France}

\date{\today}

\begin{abstract}
Noncollinear chiral spin textures in ferromagnetic multilayers are at the forefront of recent research in nano-magnetism with the promise for fast and energy-efficient devices. The recently demonstrated possibilities to stabilize such chiral structures in synthetic antiferromagnets (SAF) has raised interests as they are immune to dipolar field, hence favoring the stabilization of ultra small textures, improve mobility and avoid the transverse deflections of moving skyrmions limiting the efficiency in some foreseen applications. However, such systems with zero net magnetization are hence difficult to characterize by most of the standard techniques. Here, we report that the relevant parameters of a magnetic SAF texture, those being its period, its type (Néel or Bloch) and its chirality (clockwise or counterclockwise), can be directly determined using the circular dichroism in x-ray resonant scattering (CD-XRMS) at half integer multilayer Bragg peaks in reciprocal space. The analysis of the dependence in temperature down to 40K allows us moreover to address the question of the temperature stability of a spin spiral in a SAF sample and of the temperature scaling of the symmetric and antisymmetric exchange interactions.
 
 \end{abstract}


\maketitle %there was a space here

%\section{}
%\subsection{}
%\subsubsection{}

%\subsection{Introduction}
In condensed matter, a large variety of physical phenomena hinge on the emergence of complex chiral windings of order parameters, their observation and subsequently their control, especially in magnetism and spin-transport at the nanoscale. 
Spin-polarized scanning tunneling microscopy (SP-STM) revealed that magnetic textures with a cycloidal configuration of the magnetization and N\'eel domain walls are stabilized in ultra-thin magnetic films (one or a few atomic layers) or heavy metal layers \cite{Kubetzka,Vedmedenko, Ferrianni}. It was realized that these magnetic textures are in most cases  stabilized by interfacial Dzyaloshinskii-Moriya (DM) interaction in thin films \cite{Fert90, Heide}, the anti-symmetric analog of the Heisenberg interaction, favoring the twisting of neighboring spins around the DM vector. This interaction allows to stabilize chiral domains walls or skyrmions even at room temperature and no applied magnetic field \cite{Woo_2016, Moreau, Boulle}. However in ferromagnetic multilayers even with only a few repetitions or in single films being a few monolayer thick (and above), it has been shown that due to the presence of dipolar fields it is difficult to stabilize sub-100-nm diameter skyrmions \cite{Buttner, Legrand_2018_2} without external fields. In ferrimagnetic materials, the compensation of magnetic moments  can reduce significantly these magnetic dipolar interaction and, indeed very recently, such small magnetic skyrmions and chiral domains were reported in those systems \cite{Caretta, Streubel, Woo_2018, Hirata}. A thermally more stable alternative to rare-earth ferrimagnetic systems is to rely on synthetic antiferromagnetic (SAF) multilayers. In SAFs it has been shown that spin spirals and skyrmions  can be stabilized at room temperature  \cite{Legrand_2020} by a precise tuning of the effective perpendicular magnetic anisotropy (PMA), the DM interaction and the Ruderman–Kittel–Kasuya–Yoshida (RKKY) \cite{Parkin_90} interlayer coupling. 

In SAFs, the negligible stray field and zero total magnetization, make them extremely challenging to be investigated using techniques such as MFM, Lorentz TEM, or transmission geometry X-ray microscopy such as STXM or Ptychography. Only advanced MFM in vacuum or NV magnetometry permit clear imaging of such magnetic texture \cite{Legrand_2020, Finco}. The need for a technique allowing to probe chiral stability, 3D textures \cite{Burgos_2021} or the ultrafast dynamics \cite{Kerber_2020, Leveille_2020} of these magnetic textures is particularly crucial in the context of SAFs. Recently, we showed that the amplitude and sign of the circular dichroism in x-ray resonant magnetic scattering (XRMS) can determine the effective chirality, i.e., its type (N\'eel or Bloch) and its magnetic chirality in ferromagnetic multilayers (FM)\cite{Chauleau_2018, Legrand_2018}. In the case of multilayers in which each magnetic layer is antiferromagnetically coupled to the next, a superlattice magnetic peak corresponding to twice the charge period Bragg peak is present in reflectivity when tuning the x-ray wavelength to a core level edge of the magnetic material \cite{Tonnerre, Spezzani}. In this letter, we show that recording circular dichroism in XRMS at these positions in reciprocal space, we can access directly the chirality in SAF hosting chiral spin spirals at remanance \cite{Legrand_2020}. In addition, through temperature dependent measurement we show that these spin spirals are stable with nearly constant period in the 40-300 K range. The constant period in this temperature range indicates that the ratio of the DM and exchange interaction amplitudes is temperature independent. The nominal stacking of the SAFs prepared by sputtering deposition are listed in Table~\ref{tab:1}. They are Pt/CoFeB/Ru multilayers with three different numbers of repetitions, namely, 6, 8 and 10 (see \cite{Legrand_2020} for more details about the structures, the growth and optimization). The XRMS experiments were performed using the RESOXS diffractometer \cite{Jaouen} at the SEXTANTS beamline \cite{Sacchi_2013} of the synchrotron SOLEIL.
They were conducted in reflectivity conditions for circularly left (CL) and right (CR) incident polarizations at the Fe L$_{3}$ edge (photon energy of 707 eV), with the diffracted x-rays collected using a standard Si photodiode or a Peltier-cooled square CCD detector covering 6.1 degrees at the working distance of 26 cm in this study.

\begin{table}[h]
\caption{\label{tab:1}Structure of SAF samples investigated. The description starts from the substrate side indicated by '//', thicknesses are given in nanometers in brackets, and the numbers indexing the square brackets are the number of repetitions of the multilayer.}
\begin{ruledtabular}
\begin{tabular}{cccccc}
\hline 
No.&Multilayer stack&\\
\hline 
\hline 
I&//Pt(8)/[Co$_{0.4}$Fe$_{0.4}$B$_{0.2}$(0.9)/Ru(0.75)/Pt (0.5)]$_{6}$/Al(5)\\
II&//Pt(8)/[Co$_{0.4}$Fe$_{0.4}$B$_{0.2}$(0.9)/Ru(0.75)/Pt (0.5)]$_{8}$/Al(5)\\
III&//Pt(8)/[Co$_{0.4}$Fe$_{0.4}$B$_{0.2}$(0.9)/Ru(0.75)/Pt (0.5)]$_{10}$/Al(5)\\
\hline 
\end{tabular}
\end{ruledtabular}
\end{table}

\begin{center}
\begin{figure}[h]
 \includegraphics[width=0.5\textwidth]{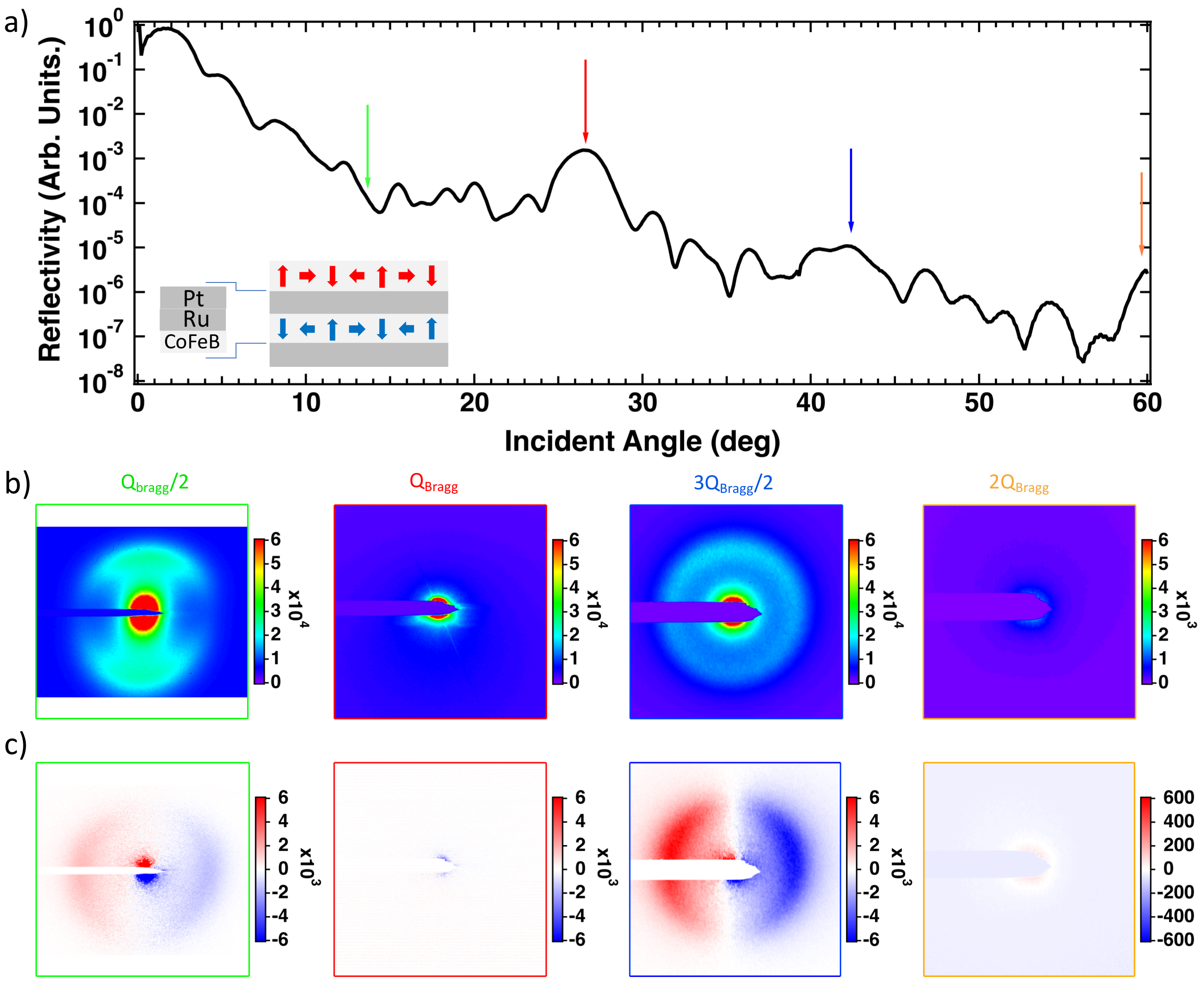}%
 \caption{(a) X-ray reflectivity recorded on Sample III at 300K and 707 eV using the sum (CL+CR) circularly polarized x-rays. The vertical red and orange arrows represent the position of the first and second order multilayer Bragg peaks arising from chemical periodicity. Green and blue arrows corresponds to chemical half Bragg peaks. (b) The sum image (CL$+$CR) using raw datas (geometrically corrected to account for the projection related to the photon incidence angle)  demonstrating a more intense magnetic scattering at half $Q_{\rm Bragg}$ and an almost vanishing one at  $Q_{\rm Bragg}$. (c) The difference image (CL$-$CR) using raw data showing intense  (about 10\% of the sum) magnetic asymmetry at $Q_{\rm Bragg}$/2 and almost vanishing one at  $Q_{\rm Bragg}$ (see text for details).\label{fig1}}
 \end{figure}
\end{center}
A typical $\theta$/2$\theta$ reflectivity, measured with the photodiode, is displayed in Fig. 1(a) for sample II. The peak corresponding to the chemical modulation of the sample period of 2.15 nm is observed around a scattering angle of $\sim$26\textsuperscript{o} labelled as $Q_{\rm Bragg}$ (see red arrow in Fig. 1(a)). The first pure magnetic peak of the AFM ordering observed in for example Ag/Ni \cite{Tonnerre} or Co/Cu\cite{Spezzani} similar antiferromagnetic multilayers is however not visible in the studied multilayers. The reason is that in case of perfectly AF coupled spin spiral with zero net magnetic moment in each layer, the pure magnetic diffraction peak is not expected. In Fig. 1(b), we display the map of the intensity sum CL + CR and in Fig. 1(c) the difference CL-CR (bottom) for four incidence angles, two corresponding to the structural order at $Q_{\rm Bragg}$ and $2Q_{\rm Bragg}$ and two for the magnetic ordering at  $Q_{\rm Bragg}/2$ and $3Q_{\rm Bragg}/2$. A diffracted ring is observed at particular angles, when the x-ray energy is tuned at Fe L$_{3}$ edge. This ring reflects the labyrinthine spin spiral periodic magnetic texture of the sample which act as a grating for x-rays. Because the magnetization in the two magnetic layers is coupled antiferromagnetically, the resonant magnetic scattering factors are reversed as well, resulting in the fact that x-ray scattering from successive planes will interfere destructively at the Bragg angle. This explains why we observe nearly no diffraction ring neither at $Q_{\rm Bragg}$ nor at 2$Q_{\rm Bragg}$. On the contrary, at $Q_{\rm Bragg}/2$ or at $3Q_{\rm Bragg}/2$, the scattered light from layers having opposite magnetization will interfere constructively and give rise to intense diffraction rings. The reasoning on the experimental map of the intensity difference (CL-CR) shown in Fig.1(c) is similar. The presence of a strong dichroic signal with a change of sign on either sides of the specular peak is a signature of the presence of a spin texture with a fixed chirality as we demonstrated for FM multilayers \cite{Chauleau_2018, Legrand_2018}.  Given the sign of the dichroic signal, we can compare it to our findings in previous studies \cite{Chauleau_2018, Legrand_2018} and conclude that counterclockwise (CCW) N\'eel magnetic textures are stabilized, as expected from the design of the multilayers with Pt underneath Co.

\begin{center}
\begin{figure}[h]
 \includegraphics[width=0.5\textwidth]{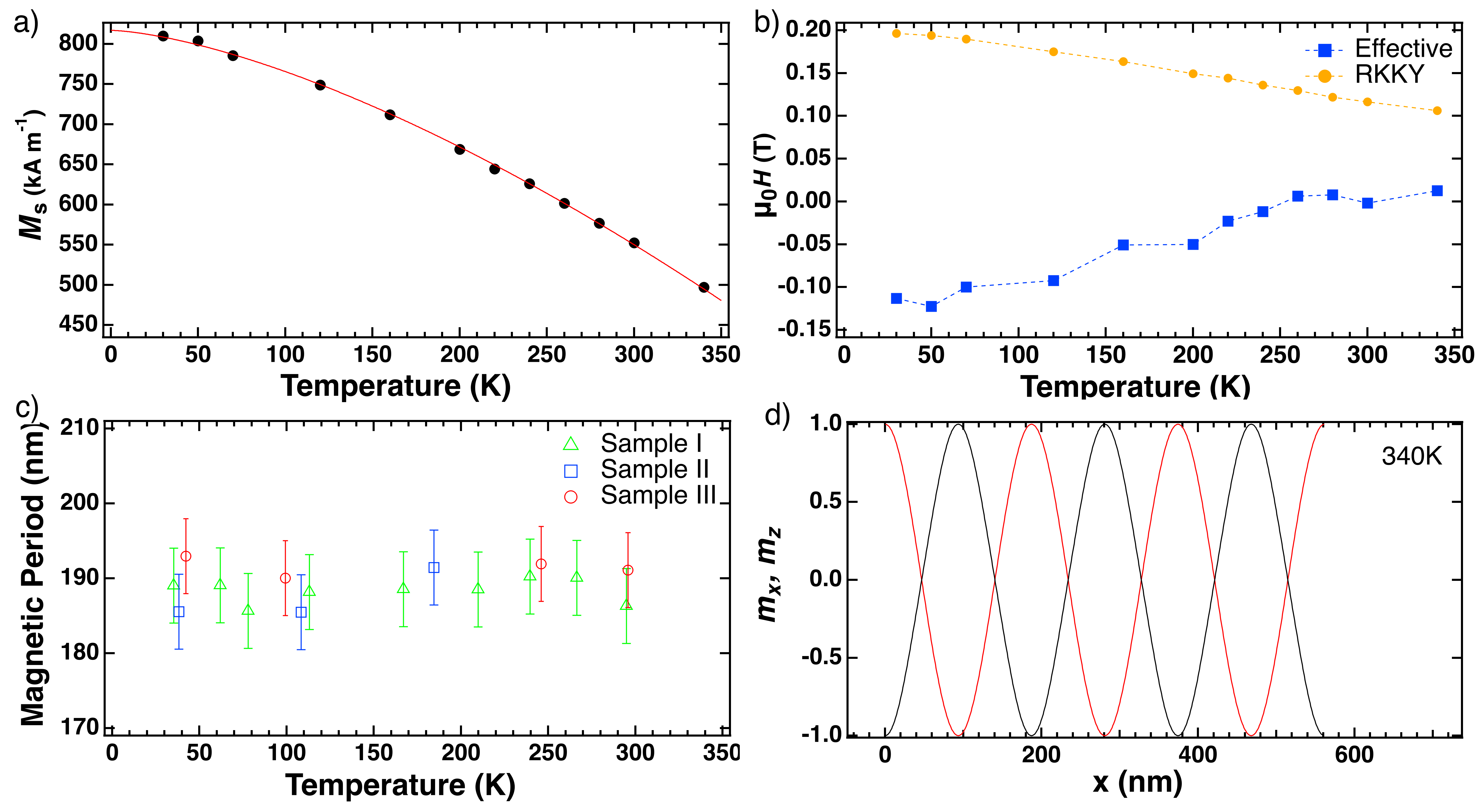}%
 \caption{(a) Temperature dependence of magnetization saturation for Sample III. (b)  Temperature dependance of $\mu_0H_{\rm eff}$ and $\mu_0H_{\rm RKKY}$ (lines are guide to the eye) for Sample III. (c) Temperature dependence of the magnetic period extract from XRMS pattern for all samples. (d) Horizontal and vertical components of the magnetization profile at 340K obtained from micromagnetic simulations using the parameters displayed in (a-c). \label{fig2}}
 \end{figure}
\end{center}

In order to analyze into more detail the intensity of the XRMS dichroism and later on its temperature dependence, we we aim to perform simulations of the XRMS signal using as input micromagnetic simulations. 
In order to determine the spatial configuration of a spin-spiral order provided by the important magnetic parameters for the simulations, we measure by SQUID the evolution of the saturation magnetization $M_{\rm s}$ as a function of temperature (See Fig. 2(a)). At room temperature the saturation magnetization is 550 kA m\textsuperscript{-1}. We also measure the temperature dependence of $\mu_0H_{\rm eff}$ and $\mu_0H_{\rm RKKY}$,  respectively the effective PMA field and the RKKY coupling field (See Fig. 2(b)). The magnetic period from the XRMS diffraction pattern is reported in Fig. 2(c). The trilayer repetition number (the sample) or the temperature are not changing the period of the spin spiral, which takes a constant value of about $190$\ nm. As the spin spiral period, $\lambda$, is proportional to the ratio of the symmetric exchange stiffness $A$ and the asymmetric (DM) one, $D$: $\lambda\approx 4\pi A/D$ \cite{Won_2005, Bode_2007, Meckler_2009}, we can conclude that the ratio between these two parameters remains constant. In order to input realistic magnetization textures to the XRMS simulation program, we use MuMax3 \cite{Vansteenkiste} self-consistent solutions, minimizing the energy for the observed period, allowing us to estimate consistent set of $A$ and $D$ parameters (see  \cite{Legrand_2020} for details). The results of the simulation are displayed in Fig. 2(d), in which the $m_x$ and $m_z$ components of the magnetization are plotted. Because $D$ is large enough $\sim$ 0.5 mJ m$^{-2}$, the $m_y $ component of $m$ is always zero. We note that the spin spiral periodicity is not changing for the three samples having different number of repetitions (see Fig. 2(c)), thus confirming the negligible dipolar field contribution in the SAF, unlike for FM multilayers \cite{Legrand_2018}.
 
\begin{center}
\begin{figure}[h]
 \includegraphics[width=0.5\textwidth]{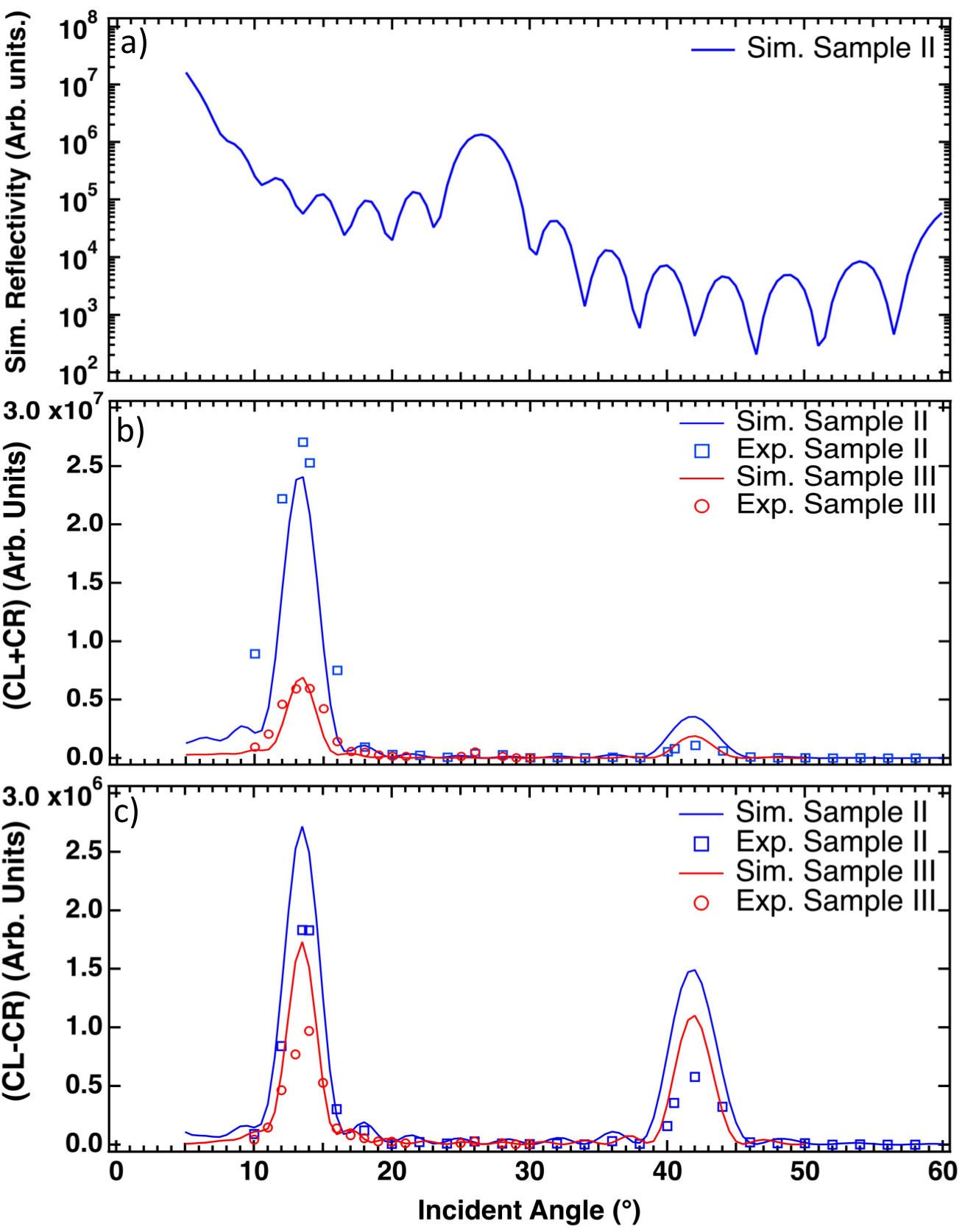}%
 \caption{(a) Simulation of reflectivity for sample II as a function of angle (b) Simulations (lines) and experimental (dots) sum (CL$+$CR) as a function of incident angle for sample II (blue) and III (red). (c) Simulations lines) and experimental (dots) sum (CL$-$CR) as a function of angle for sample II (blue) and III (red). Both simulation and experiment have been done for x-ray energy of 707eV. (See text for details) \label{fig3}}
 \end{figure}
\end{center}

Based on the results of the micromagnetic simulations, we now simulate the expected specular reflectivities and XRMS patterns and compare them with the experiments. While we have already simulated XRMS for FM multilayers assuming that all the magnetic layers have the same magnetization and considering only one layer \cite{Chauleau_2018, Leveille_2020},  this is not a valid hypothesis in the case of AFM coupled multilayers. As a consequence, two layers have to be considered for the magnetic unit cell. 
Unlike for hard X-ray range, for which kinematical approximations can be used to simulate XRMS \cite{Seve_1999, Ishimatsu_1999, Jaouen_2002, Nelson_1999}, simulation of resonant magnetic reflectivity require dynamical approximation as  the classical description with Maxwell equations and a permittivity built from the quantum scattering amplitude \cite{Elzo_2012, Macke_2014}. This approach has been successfully used in complex oxide heterostructures \cite{Gibert_2016, Fabbris_2018} demonstrating that the strong absorption and multiple scattering that become important especially at core level resonance of 3$\it{d}$ metals are well taken into account. Therefore, for this study we use the distorted wave born approximation (DWBA) to simulate XRMS \cite{Lee_2003, Lee_2003bis,Flewett_2019, Flewett_2020}, and we analyze the reflectivity simulated at Fe L$_{3}$ edge (707 eV) for the sample II and III structure described in Table 1. The simulated reflectivity curve in Fig. 3(a) exhibits a 1$^{st}$ and 2$^{nd}$ Bragg peaks respectively around $\sim$26\textsuperscript{o} and $\sim$60\textsuperscript{o} similarly to the experiment (Fig1. (a)).  Even if we observe several differences, mainly because we used an idealized roughness-free sample in our simulation where only magnetic/non-magnetic interfaces were considered, it does not alterate the main point of this discussion.  In Fig. 3(b), we display using open symbols the sum signal (CL+CR) angular dependence experimentally obtained by radial integration of the magnetic diffraction for both left and right circular polarization for sample II and III as well as the simulated ones using solid lines. CL and CR experimental intensities have been evaluated after removal of the diffuse background coming from the specular following the same approach previously used in L\'eveill\'e  $ \it{et}$  $\it{al.}$ \cite{Leveille_2020}

We find in the simulated XRMS that the (CL$+$CR) intensity is almost vanishing at $Q_{\rm Bragg}$ and 2$Q_{\rm Bragg}$ positions but is maximal at $Q_{\rm Bragg}$/2 and $3Q_{\rm Bragg}/2$. In Fig. 3(c), we plot the experimental dichroism (CL$-$CR) with symbols and the corresponding simulations with lines. For both samples and for (CL$+$CR) and (CL$-$CR) a very good agreement around $Q_{\rm Bragg}$/2 is found. For 3$Q_{\rm Bragg}$/2 the simulations overestimate the sum and dichroism intensities. This difference can be explained as assuming an ideal structure for the simulation overestimates constructive interference at the Bragg angles. Again the impact of the roughness of the multilayers is not taken into account in the simulations, which also leads to an over estimation of scattering intensity at high scattering angles. 

\begin{center}
\begin{figure}[h]
 \includegraphics[width=0.5\textwidth]{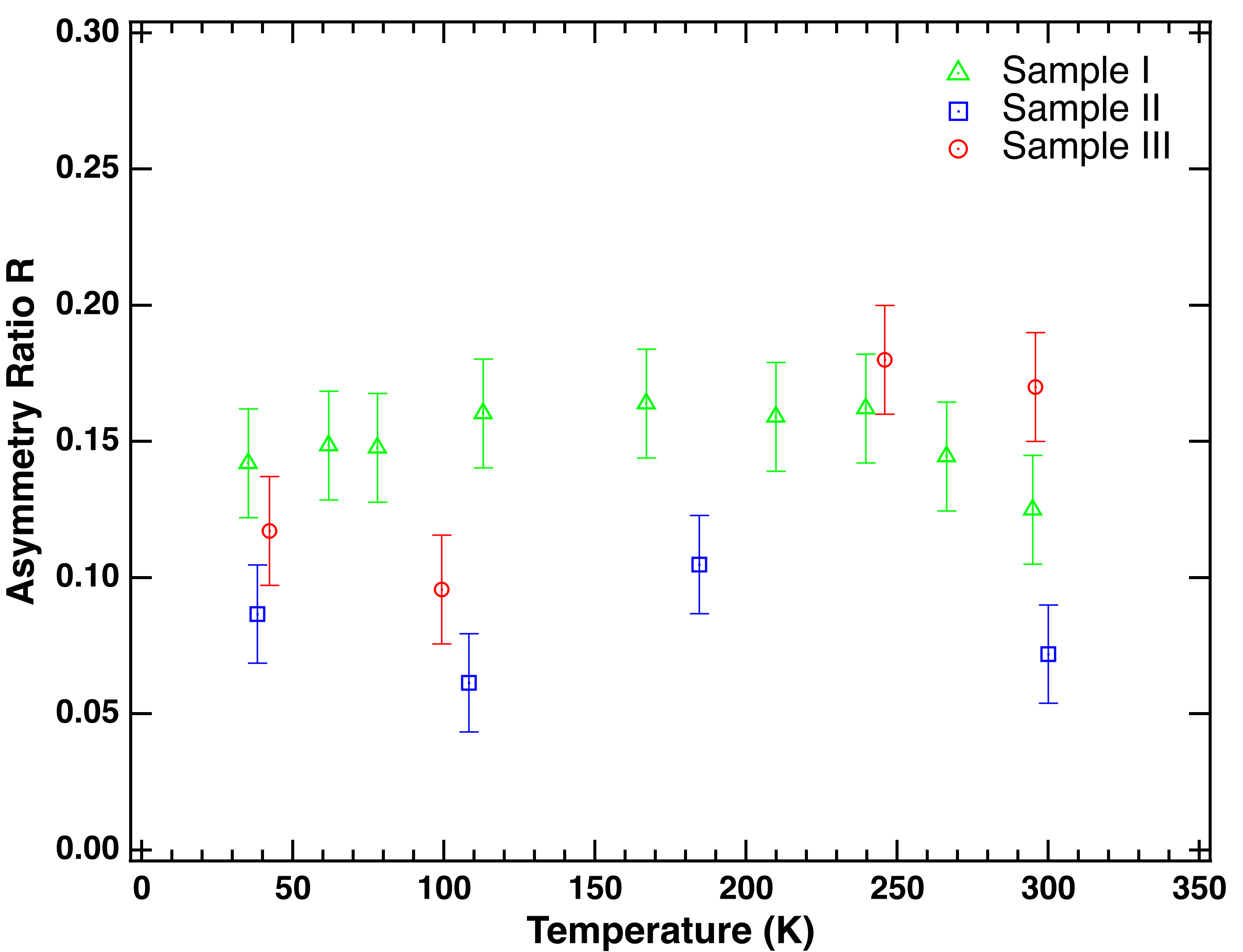}%
 \caption{Temperature dependence of the asymmetry ratio for Sample I (red dots), Sample II (green dots) and Sample III, (blue dots) measured at $Q_{\rm Bragg}$/2. \label{fig4}}
 \end{figure}
\end{center}

We now turn to the temperature evolution of the magnetic asymmetry ratio, defined as R=(CL-CR)/(CL+CR), measured at  $Q_{\rm Bragg}$/2. To this aim, we cooled down the samples to $40$\,K (lowest reachable temperature in the experiment) and measured the dichroism whilst gradually increasing the temperature. In Fig. 4, we see that despite the noticeable change of saturation magnetization, $\mu_0H_{\rm eff}$ and $\mu_0H_{\rm RKKY}$, the spin spiral state (period and chirality) is a metastable state that survives to changes of the magnetic parameters over several tens of percent of relative variations. This demonstrates that the spin spiral state can be stabilized for a large range of magnetic parameters and that it is most probably possible to stabilize them for a rich variety of material compositions. Moreover, we have seen that the period of the spin-spiral order $4\pi A/D$ remains constant in the probed temperature range of 40-300K, within which Ms varies by 30$\%$. This actually provides strong evidence for one specific scenario of temperature scaling for the DM interaction. According to mean field theory and its refinements, the symmetric (Heisenberg) exchange can be described by an amplitude $A$ evolving with temperature as a function of $M_{\rm s}$, following an $M_{\rm s}(T)^{\sim2}$ law, while $M_{\rm s}$ follows Bloch law $M_{\rm s}(T)$=$M_{\rm s}(0)$(1-($T$/$T_{\rm c}$)$^{(3/2)}$). We can thus deduce from the constant ratio $A$/$D$ that $D$ decreases with temperature following a similar $M_{\rm s}(T)^{\sim2}$ law, in agreement with other works \cite{Zhou_2020} .

In summary, we report on the possibility of using circular dichroism in XRMS to determine the chiral properties of magnetic texture in SAF multilayers. We show experimentally  that the magnetic diffraction and the difference between left and right circular polarization are maximal at positions in reciprocal space corresponding to the doubling of the chemical period, i.e. equal to the magnetic period in the sample. Our experimental findings have been confirmed by XRMS simulation using the DWBA  with the results of micromagnetic simulations for input. We found an excellent agreement confirming our experimental result for both the diffracted intensity (CL+CR) and the dichroism (CL-CR). Finally, we also show that independent of the number of periods, unlike for the FM multilayers, the period and the chirality of the spin spiral remain constant over a broad temperature range indicating a constant ratio  $A$/$D$ and thus their similar temperature scaling.

In a broader perspective, circular dichroism in x-ray scattering appears to be an unique tool to study chiral magnetic texture in SAF materials. We also would like to point out, that this study opens the way for time resolved studies of magnetic texture in AFM as it has recently been done for FM multilayers using the femtosecond X-rays pulses available at x-ray free electron lasers (XFEL)\cite{Leveille_2020, Kerber_2020} or high harmonic generation (HHG) sources \cite{Vodungbo_2012, Fan_2020}.

\subsection{Acknowledgments}
Financial supports from the French National Research Agency (ANR) with TOPSKY (ANR-17-CE24-0025), FLAG-ERA SographMEM (ANR-15-GRFL-0005, PCI2019-111908-2) and as part of the “Investissements d'Avenir” program SPiCY (ANR-10-LABX-0035), from Mineco grant AEI/FEDER, UE (FIS2016-7859-C3-2-R) and from the Horizon2020 Framework Program of the European Commission under FET-Proactive Grant SKYTOP (824123) are acknowledged.

%\bibliography{TR-XRMR_Ni_biblio}

\end{document}